\documentclass[%
reprint,
superscriptaddress,
amsmath,
amssymb,
aps,
pra,
longbibliography
]{revtex4-1}

\usepackage{graphicx}
\usepackage{dcolumn}
\usepackage{bm}
\usepackage[T1]{fontenc}
\usepackage[utf8]{inputenc}
\usepackage{times}
\usepackage{amsmath}
\usepackage{amsmath,amsfonts,amssymb,mathtools}
\usepackage{hyperref}
\hypersetup{hypertex=true,
            colorlinks=true,
            linkcolor=red,
            anchorcolor=blue,
            citecolor=blue,
          unicode=true, breaklinks=true}
              \usepackage{subfigure}
\usepackage{bbold}
\usepackage{tensor}

\begin{document}

\preprint{APS/123-QED}

\title{Controllable generation and spatial phase-modulation of vortex beams in cascade-type atomic ensembles}%
\author{Yueqian Li}
\affiliation{Center for Quantum Sciences and School of Physics, Northeast Normal University, Changchun 130024, China}
\author{Yan Zhang}
\email{zhangy345@nenu.edu.cn}
\affiliation{Center for Quantum Sciences and School of Physics, Northeast Normal University, Changchun 130024, China}


\date{\today}

\begin{abstract}
We propose a four-level cascade-type atomic system in which two reverse four-wave mixing (FWM) processes can coexist via optical transitions in distinct pathways. This enables the generation of high- or low-frequency vortex beams on demand and facilitates spatial phase modulation of the vortex beam intensity distribution. 
The two reverse FWM processes can yield the sum-frequency vortex beam with high frequency and the difference-frequency vortex beam with low frequency, corresponding to frequency up-conversion and down-conversion, respectively. 
The closed-loop configuration's phase sensitivity allows for asymmetric phase modulation of the spatial distributions of the two output vortex beams. 
Additionally, increasing the phase enables spatial rotation of the output beams. Our proposed scheme holds promise for potential applications in quantum nonlinear optics and vortex exchange.
\end{abstract}

\maketitle

\section{Introduction}
As an important way to modulate the photons and light, electromagnetically induced transparency (EIT) has attracted people's attention \cite{1,2,3}.
EIT media with suppressed absorption and anomalous dispersion can realize optical nonlinear effects such as slow and fast light \cite{4,5}, optical storage and retrieval \cite{6,7}, etc. Recently, more and more attention has been paid to the application of special beams in the regime of EIT, especially optical vortices \cite{2,9,10,11,12}. 

The vortex beam, whose intensity is zero in the center and is distributed in a ring along the propagation direction, is a special light field with the helical wavefront phase as a function of azimuth coordinates and orbital angular momentum (OAM) \cite{13}. 
Its OAM provides additional possibilities for coherent control of light\cite{14,15,16}. 
Notably, the additional degrees of freedom suggest that vortex beams have more applications for quantum computing, quantum teleportation, and quantum information storage \cite{17}. 
The application of vortex light fields in the field of optical micro-control has led to a large number of studies of OAM in light fields, especially Laguerre-Gaussian (L-G) light field \cite{LG1, LG2} as a typical vortex beam with a field distribution. 
All types of vortex beams can be simply regarded as a linear superposition of L-G modes. 
In studies of the three-level EIT structure, due to that, the zero center field of the strong vortex control beam destroys the EIT and leads to significant loss; only the probe field can be considered with the vortices \cite{13,18}.
Thus, by extending the level structure and introducing an additional control field without vortices, the OAM transfer from the vortex control field can be observed when the probe and the vortex control fields are opposite \cite{20,21}. 
Moreover, the exchange of OAM can be realized using the noise-induced coherence effect \cite{2022PRAHmedi}. 
These demonstrations of the easy adjustment of the propagation properties of optical media by the vortex beams make them an effective tool, usually involving the four-wave mixing (FWM) process.

FWM is a typical high-order nonlinear effect in multilevel systems, which results from the third-order nonlinear polarization of the medium \cite{23,25,26}. 
The FWM phenomenon based on EIT has attracted much attention due to the enhancement of the nonlinear polarizability and the inhibition of the linear polarizability of media in the regime of EIT, such as the first experimental implementation of compressed light based on the FWM effect using the EIT atomic ensemble\cite{squeeze}, FWM generation with enhanced efficiency \cite{FWMa1, FWMa2, FWMa3, FWMa4}. 
With the possible applications in nonlinear optics, much attention has been paid to exploring controllable modulation of input signal resulting from the interaction of the generated FWM field with the incident field \cite{2018PRAZub}.

In this paper, we propose a four-level cascade-type structure with two reverse FWM processes for controllable generation and spatial modulation of vortex beams. 
In this system, we can realize the frequency-up-conversion or frequency-down-conversion generation of the vortex beam. 
The OMA can switch between the high-frequency field and the low-frequency field. 
It is found that the light intensity of the transmitted probe field is affected by the relative phase.

\section{Model and Equations}\label{SecII}

\begin{figure}[t]
\includegraphics[width=0.48\textwidth]{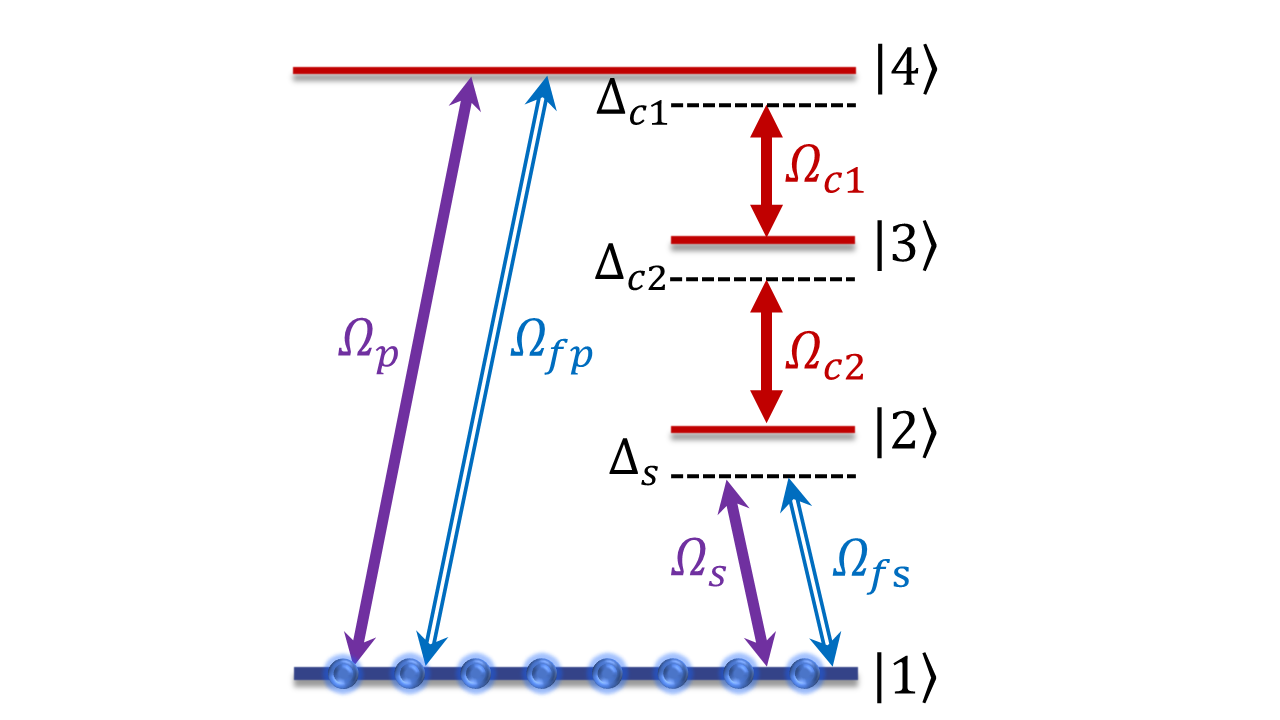}
\caption{ The sketch of the four-energy cascade-type atomic system interacting with two weak probe fields $\Omega_{p,s}$, and two strong coupling fields $\Omega_{c1,c2}$. 
Two FWM processes in this closed-loop system generate the FWM sum-frequency field $\Omega_{fp}$ and the FWM difference-frequency field $\Omega_{fs}$.}
\label{Fig1}
\end{figure}

As shown in Fig.~\ref{Fig1}, We consider a close-loop four-level atomic system in the cascade configuration. Two weak incident probe and signal fields are with the Rabi frequencies $\Omega_{p,s}$ and the wave-vector $\mathbf{k}_{p,s}$, respectively, while the strong coupling fields with $\Omega_{c1,c2}$, respectively. These fields, which may carry OAM on demand, couple the corresponding atomic transitions, respectively, as shown in Fig.~\ref{Fig1}. 
Such a setup can be realized experimentally. For example, we can use the $^{87}Rb$ cold atomic ensemble confined in a magneto-optical trap to form the cascade-type atomic scheme. 
The four energy levels $|1\rangle$, $|2\rangle$, $|3\rangle$ and $|4\rangle$ can correspond to the states $5S_{1/2} (F=2)$, $5P_{1/2}$, $5D_{3/2}$, $nP_{3/2}$ ($n>10$). 
The frequency of transition $| 2\rangle\rightarrow | 1\rangle$ is $795$ nm, the frequency of $| 3\big \rangle\rightarrow | 2 \rangle$  is $762$ nm, and the frequency of $ | 4 \rangle\rightarrow | 3 \rangle$ is $1.3-1.5$ $\mu$m.

The atom-field system forms a closed-loop four-level structure with two co-exist FWM transition paths: $ |1\rangle \rightarrow  |2\rangle \rightarrow  |3 \rangle \rightarrow  |4 \rangle \rightarrow  |1 \rangle$ and its inverse process. 
The first (second) one, being the frequency up-conversion (down-conversion), corresponds to the sum-frequency (difference-frequency) generation process to generate the field with Rabi frequency $\Omega_{fp}$ ($\Omega_{fs}$). 
Two FWM-generated fields meet the phase matching conditions $\Delta_\mathbf{k}=\mathbf{k}_{fs}-\mathbf{k}_{p}-\mathbf{k}_{c1}-\mathbf{k}_{c2}=\mathbf{k}_{fp}-\mathbf{k}_{s}-\mathbf{k}_{c1}-\mathbf{k}_{c2}=0$. The system Hamiltonian is
\begin{equation}
\begin{split}
H=&\sum_{j=1}^4\omega_j\big |j\big \rangle\big \langle j\big | \\
&-\frac{1}{2} \left ( \Omega_s e^{-i\omega_s t}+\Omega_{fs} e^{-i\omega_{fs}t}\right) \big |2\big \rangle \big \langle 1\big | \\
&-\frac{1}{2} \Omega_{c2} e^{-i\omega_{c2} t}\big |3\big \rangle \big \langle 2 \big | -\frac{1}{2} \Omega_{c1} e^{-i\omega_{c1} t} \big |4\big \rangle \big \langle 3 \big | \\
&-\frac{1}{2} \left ( \Omega_p e^{-i\omega_p t}
+\Omega_{fp} e^{-i\omega_{fp} t} \right)\big |4\big \rangle \big \langle 1 \big |  + \text{H.c.} ,
\label{Eq_H1}
\end{split}
\end{equation}
where $\omega_{s, fs, c2, c1, p, fp}$ represent the frequencies of the corresponding fields while $\omega_j$ denotes the intrinsic frequency of the corresponding levels, respectively. 
The frequency of the generated sum-frequency and difference-frequency fields meet the requirements of $\omega_{fs}$=$\omega_p$-$\omega_{c1}$-$\omega_{c2}$, $\omega_{fp}$=$\omega_s$+$\omega_{c2} $+$\omega_{c1}$. Under the interaction representation, the Hamiltonian becomes 
\begin{equation}
\begin{split}
\label{eq2}
H_I=&\Delta_s |2 \rangle  \langle 2 |+\left(\Delta_s+\Delta_{c2}\right) |3 \rangle  \langle 3  |\\
&+(\Delta_s+\Delta_{c2}+\Delta_{c1})|4  \rangle  \langle 4  |\\
&-\frac{1}{2}(\Omega_s+\Omega_{fs})|2\rangle  \langle 1|-\frac{1}{2}\Omega_{c2} |3 \rangle  \langle 2 |  \\
&-\frac{1}{2}\Omega_{c1} |4\rangle \langle 3 |-\frac{1}{2}(\Omega_{p} +\Omega_{fp}) |4 \rangle  \langle 1 |+\text{H.c.}.
\end{split}
\end{equation}
Here $\Delta_s=\omega_{21}-\omega_s$, $\Delta_{c2}=\omega_{32}-\omega_{c2}$, $\Delta_{c1}=\omega_{43}-\omega_{c1}$
denote the frequency detunings between the probe, drive, and control fields and the corresponding transitions, respectively, where $\omega_{21,32,43}$ denote the transition frequencies of $|2 \rangle \leftrightarrow |1 \rangle$, $|3 \rangle \leftrightarrow |2 \rangle$, and $|4 \rangle \leftrightarrow |3 \rangle$, respectively.

Using the dynamic evolution master equation, we can obtain the non-diagonal density matrix elements as 
\begin{equation}
\begin{split}
\dot{\rho_{21}}=&-\Gamma_{21}\rho_{21}+\frac{i}{2}(\Omega_s+\Omega_{fs})(\rho_{11}-\rho_{22})\\
&+\frac{i}{2}\Omega_{c2}^*\rho_{31}-\frac{i}{2}(\Omega_p+\Omega_{fp})\rho_{24},\\
\dot{\rho_{31}}=&-\Gamma_{31} \rho_{31}+\frac{i}{2}\Omega_c2\rho_{21}+\frac{i}{2}\Omega_{c1}^*\rho_{41}\\
&-\frac{i}{2}(\Omega_s+\Omega_{fs})\rho_{32}-\frac{i}{2}(\Omega_p+\Omega_{fp})\rho_{34},\\
\dot{\rho_{32}}=&-\Gamma_{32}\rho_{32}+\frac{i}{2}\Omega_{c2}(\rho_{33}-\rho_{44})\\
&+\frac{i}{2}\Omega_{c1}^*\rho_{42}-\frac{i}{2}(\Omega_s^*+\Omega_{fs}^*)\rho_{31},\\
\dot{\rho_{41}}=&-\Gamma_{41}\rho_{41}+\frac{i}{2}\Omega_{c1}\rho_{31}\\
&+\frac{i}{2}(\Omega_p+\Omega_{fp})(\rho_{11}-\rho_{44})-\frac{i}{2}(\Omega_s+\Omega_{fs})\rho_{42},\\
\dot{\rho_{42}}=&-\Gamma_{42}\rho_{42}+\frac{i}{2}\Omega_{c1}\rho_{32}+\frac{i}{2}\Omega_{c2}\rho_{43},\\
&+\frac{i}{2}(\Omega_p+\Omega_{fp})\rho_{12}-\frac{i}{2}(\Omega_s^*+\Omega_{fp}^*)\rho_{41},\\
\dot{\rho_{43}}=&-\Gamma_{43}\rho_{43}+\frac{i}{2}\Omega_{c1}(\rho_{33}-\rho_{44})\\
&-\frac{i}{2}\Omega_{c2}^*\rho_{42}+\frac{i}{2}(\Omega_p+\Omega_{fp})\rho_{13},\\
\end{split}
\end{equation}
with $\Gamma_{21}=\gamma_{21}+i\Delta_s$, $\Gamma_{31}=\gamma_{31}+i(\Delta_s+\Delta_{c2})$, $\Gamma_{32}=\gamma_{32}+i\Delta_{c2}$, $\Gamma_{41}=\gamma_{41}+i(\Delta_s+\Delta_{c2}+\Delta_{c1})$, 
$\Gamma_{42}=\gamma_{42}+i(\Delta_{c2}+\Delta_{c1})$, $\Gamma_{43}=\gamma_{43}+i\Delta_{c1}$, where $\gamma_{ij}$ is the coherence decay rate. Here, $\gamma_{41}=\gamma_{21}=\gamma$. 

Under the weak field approximation condition and considering the final population to be in the ground state $|1\rangle$, we can obtain the first-order steady-state solutions of the non-diagonal elements for the generated fields as
\begin{equation}
\begin{split}
\label{eq9}
&\rho_{21}^{(1)}=\frac{i(\Omega_s+\Omega_{fs})}{2\Gamma_{21}},\\ 
&\rho_{41}^{(1)}=\frac{i\Gamma_{31}(\Omega_p+\Omega_{fp})}{2\xi},\\ 
\end{split}
\end{equation}
and third steady-state solutions as
\begin{equation}
\begin{split}
\label{eq11}
&\rho_{21}^{(3)}=\frac{i^3\Omega_p\Omega_{c1}^*\Omega_{c2}^*}{8\Gamma_{21}\xi}+\frac{i^3\Omega_{fp}\Omega_{c1}^*\Omega_{c2}^*}{8\Gamma_{21}\xi},\\
&\rho_{42}^{(3)}=\frac{i^3\Omega_s\Omega_{c2}^*\Omega_{c1}^*}{8\Gamma_{21}\xi}+\frac{i^3\Omega_{fs}\Omega_{c2}^*\Omega_{c1}^*}{8\Gamma_{21}\xi}, 
\end{split}
\end{equation}
with $\xi=\Gamma_{31}\Gamma_{41}+\big |\Omega_{c1}\big |^2/4$.
The first-order solutions generally affect the dispersion and absorption properties and the resulting FWM-generated fields. 
The third-order solutions exhibit the third-order nonlinearity, where the first term describes the conversion to the sum frequency (difference frequency) field in the FWM process, and the second term shows the backward nonlinear process of the FWM-generated field.

For dynamic generation and evolution for the probe and signal fields, we combine Eqs.~(\ref{eq9}) and (\ref{eq11}) with Maxwell wave equations, which can ignore transverse second-order partial derivations in the slowly varying envelope approximation. In addition, our interest is in the atomic response to the long probe and signal pulses, such that $(1/c)\partial\Omega_{p,s}/\partial t=0$. 
Then, the Maxwell wave equations are simplified to \cite{21,2023PRAYang}
\begin{equation}
\begin{split}
\label{eq15}
\frac{\partial\Omega_{fp}}{\partial z}&=i\kappa_{14} \left [\frac{i\Gamma_{31}\Omega_{fp}}{2\xi_{21}}+\frac{i^3\Omega_{s}\Omega^*_{c2}\Omega^*_{c1}}{8\Gamma_{21}\xi} \right ],\\
\frac{\partial\Omega_{fs}}{\partial z}&=i\kappa_{12}\left [\frac{i\Omega_{fs}}{2\Gamma_{21}}+\frac{i^3\Omega_{p}\Omega_{c1}^*\Omega_{c2}^*}{8\Gamma_{21}\xi} \right ],\\ 
\frac{\partial\Omega_{p}}{\partial z}&=i\kappa_{14} \left [\frac{i\Gamma_{31}\Omega_{p}}{2\xi}+\frac{i^3\Omega_{fs}\Omega_{c2}\Omega_{c1}}{8\Gamma_{21}\xi} \right ],\\
\frac{\partial\Omega_s}{\partial z}&=i\kappa_{12}\left [ \frac{i\Omega_s}{2\Gamma_{21}}+\frac{i^3\Omega_{fp}\Omega^*_{c1}\Omega^*_{c2}}{8\Gamma_{21}\xi} \right ].
\end{split}
\end{equation}

Setting $\Omega_{i0}$ and $\phi_i$ ($i=p,s,c1,c2$) are the initial Rabi frequency and optical phase for the relevant fields with $z=0$, respectively, such a closed-loop system is sensitive to the relative phase $\phi=\phi_s-\phi_p+\phi_{c1}+\phi_{c2}$. 
Then, we obtain the FWM-generated fields as
\begin{align}
\label{eq19}
\Omega_{fp}=\frac{c}{\lambda} \Omega_{s_0} e^{-i\phi} \left ( \zeta_1-\zeta_2 \right ), \\ 
\Omega_{fs}=\frac{\kappa c}{\lambda} \Omega_{p_0} e^{i\phi}  \left ( \zeta_1-\zeta_2 \right ),
\end{align}
and the evolving probe fields as 
\begin{equation}
\begin{split}
\Omega_s&=\Omega_{s_0}  \left(\frac{a-\kappa b+\lambda}{2 \lambda}\zeta_2 -\frac{a-\kappa b-\lambda}{2 \lambda}\zeta_1 \right ),\\
\Omega_p&=\Omega_{p_0} \left( \frac{a-\kappa b+\lambda}{2 \lambda}\zeta_1  - \frac{a-\kappa b-\lambda}{2 \lambda}\zeta_2 \right ) ,\\
\end{split}
\end{equation}
with $\kappa=\kappa_{12}/\kappa_{14}$, $\lambda$=$\sqrt{(\kappa b-a)^2+4\kappa |c|^2}$, $a=-\Gamma_{31}/2\xi$, $b=-1/2\Gamma_{21}$, $c=\Omega_{c2}\Omega_{c1}/8\Gamma_{21}\xi$, $\zeta_1=e^{(a+\kappa b+\lambda)Z/2}$, $\zeta_2=e^{(a+\kappa b-\lambda)Z/2}$, and the effective propagation distance $Z=\kappa_{03} z$.
The four fields could be the L-G vortex beam having the OAM $\hbar l_i$ with the topological charge $l_i$ along the propagation axis $z$. 
Then, the initial field can be characterized by $  \Omega_{i0}=\varepsilon_{i0}(\frac{r}{w})^{\lvert l_i \rvert } e^{-r^2/w^2} e^{il \Phi}$, where $r$, $w$, $\Phi$, and $\varepsilon_{0}$ are the vortex beam radius, the beam waist, azimuthal angle, and the strength of the beam corresponding to the initial case, respectively.

\section{Results and discussion}

\begin{figure}[t]
\centering
\includegraphics[width=0.45\textwidth]{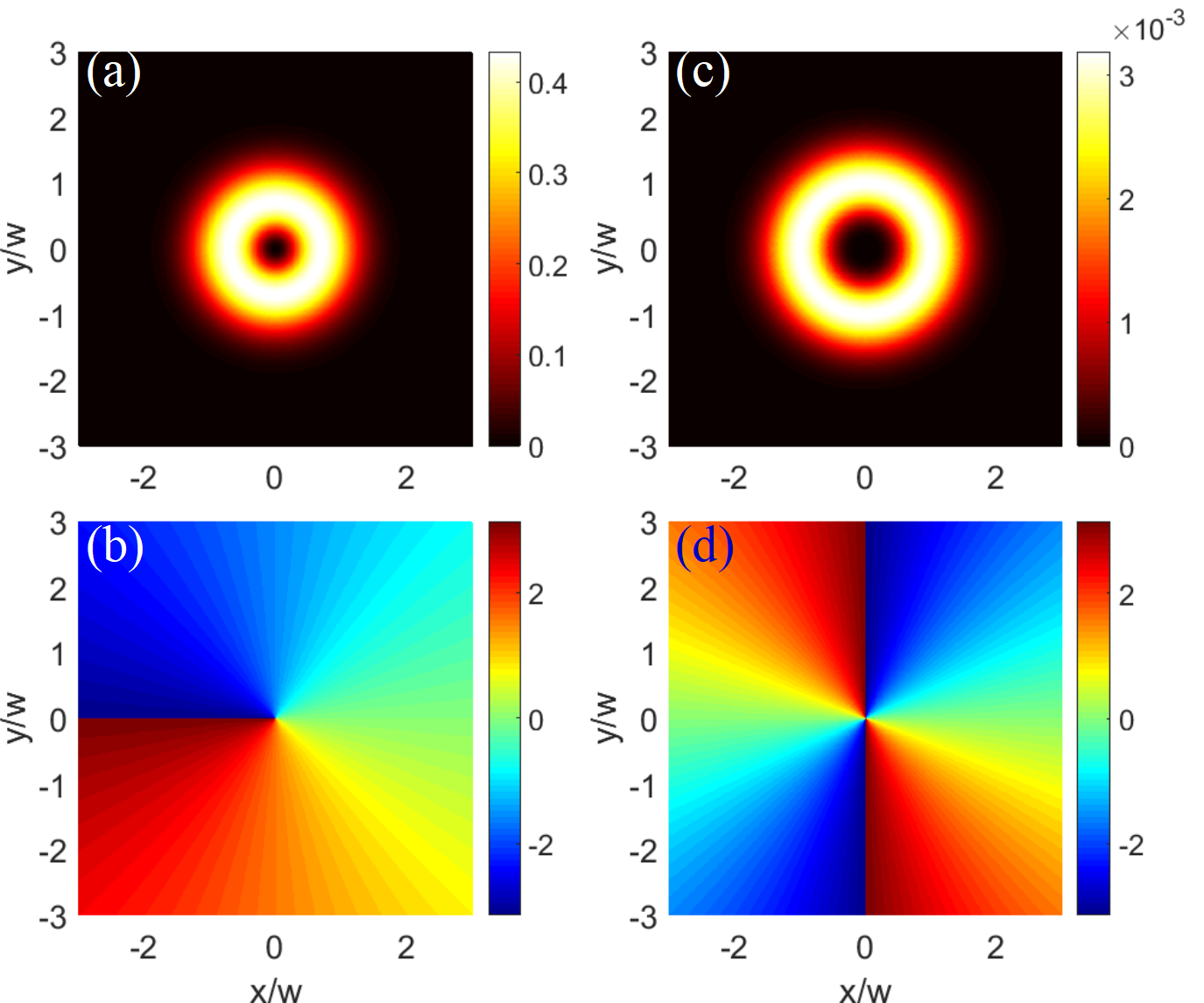}
\caption{(a) Intensity $\left |\Omega_{fp}\right |^2/\gamma^2$ distribution on the plane of the axis $x$ and $y$ and (c) phase of sum-frequency field $\Omega_{fp}$ with $\Omega_{p0}=0$ at the effective propagation distance $Z=10\gamma$ when the probe field $\Omega_s$ with OAM number $l_s=1$ and $\varepsilon_{s0}= 0.1\gamma$. 
(b) Intensity $\left |\Omega_{fs}\right |^2/\gamma^2$ distribution and (d) phase of sum-frequency field $\Omega_{fs}$ with $\Omega_{s0}=0$ at the effective propagation distance $Z=10\gamma$ when the probe field $\Omega_p$ with OAM number $l_p=2$ and $\varepsilon_{p0}= 0.1\gamma$.
(e) Intensity of generated field $\Omega_{fs} $ versus the dimensionless distance $r/w$ from the vortex core for $l=1,2,3$.
The other parameters are $\phi=0$, $\Omega_{c1}=\Omega_{c2}=10\gamma$ 
$\gamma_{31}=0.05\gamma$, $\Delta_{c1}=\Delta_{c2}=\Delta_s=0$, 
and $\kappa=0.1$.} 
\label{fig2}
\end{figure}

Here, we examine the system's ability to transfer the optical vortex to the FWM-generated sum-frequency (difference-frequency) field $\Omega_{fp}$ ($\Omega_{fs}$) from the incident field $\Omega_{s}$ ($\Omega_{p}$). 
When the initial field $\Omega_s$ or $\Omega_p$ is a vortex beam carrying OAM, we can obtain
\begin{equation}
\begin{split}
\label{eq22}
\Omega_{fp}=\frac{\kappa c}{\lambda}\varepsilon_{s_0}(\frac{r}{w})^{\lvert l_s \rvert }e^{-\frac{r^2}{w^2}} \left ( \zeta_1-\zeta_2 \right ) e^{i (l_j \Phi-\phi)},
\end{split}
\end{equation}
or
\begin{equation}
\begin{split}
\Omega_{fs}=\frac{c}{\lambda}\varepsilon_{p_0}(\frac{r}{w})^{\lvert l_p \rvert } e^{-\frac{r^2}{w^2}} (\zeta_1-\zeta_2) e^{i (l_j \Phi+\phi)}.\\
\end{split}
\end{equation}

First, we only input one of the probe fields as the low-frequency $\Omega_{s}$ with OAM number $l_{s}=1$. As shown in Fig.~\ref{fig2}(a), under the resonance condition, the sum-frequency vortex beam $\Omega_{fp}$ with a toroidal transmission profile is generated from the FWM process due to the input of $\Omega_{s}$. As shown in Fig.~\ref{fig2}(b), the spiral phase distribution for vortex beam $\Omega_{fp}$ has a phase transition from 2$\pi$ to -2$\pi$, i.e., the number of the phase singularity being $1$. This refers to the OAM number $l_{fp}=1$ \cite{2023PRAYang}, meaning that the OAM of the initial field can be totally transferred to the generated field. Then, when only inputting one of the probe fields as the high-frequency $\Omega_{p}$ with OAM number $l_{p}=2$, as shown in Fig.~\ref{fig2}(b) and (d), the difference-frequency vortex beam $\Omega_{fs}$ with OAM number $l_{fs}=2$ is generated from the other FWM process.
Thus, using such a double-FWM system, we can realize the high-frequency (such as ultraviolet light) vortex beam by inputting low-frequency light or the reverse process on demand. Moreover, the OAM can be perfectly exchanged between the beams with different frequencies. In addition, the radius of the vortex beam increases as the value of the orbital angular momentum $l$ increases, and the intensity increases as well. 


\begin{figure*}[t]
\includegraphics[width=0.7\textwidth]{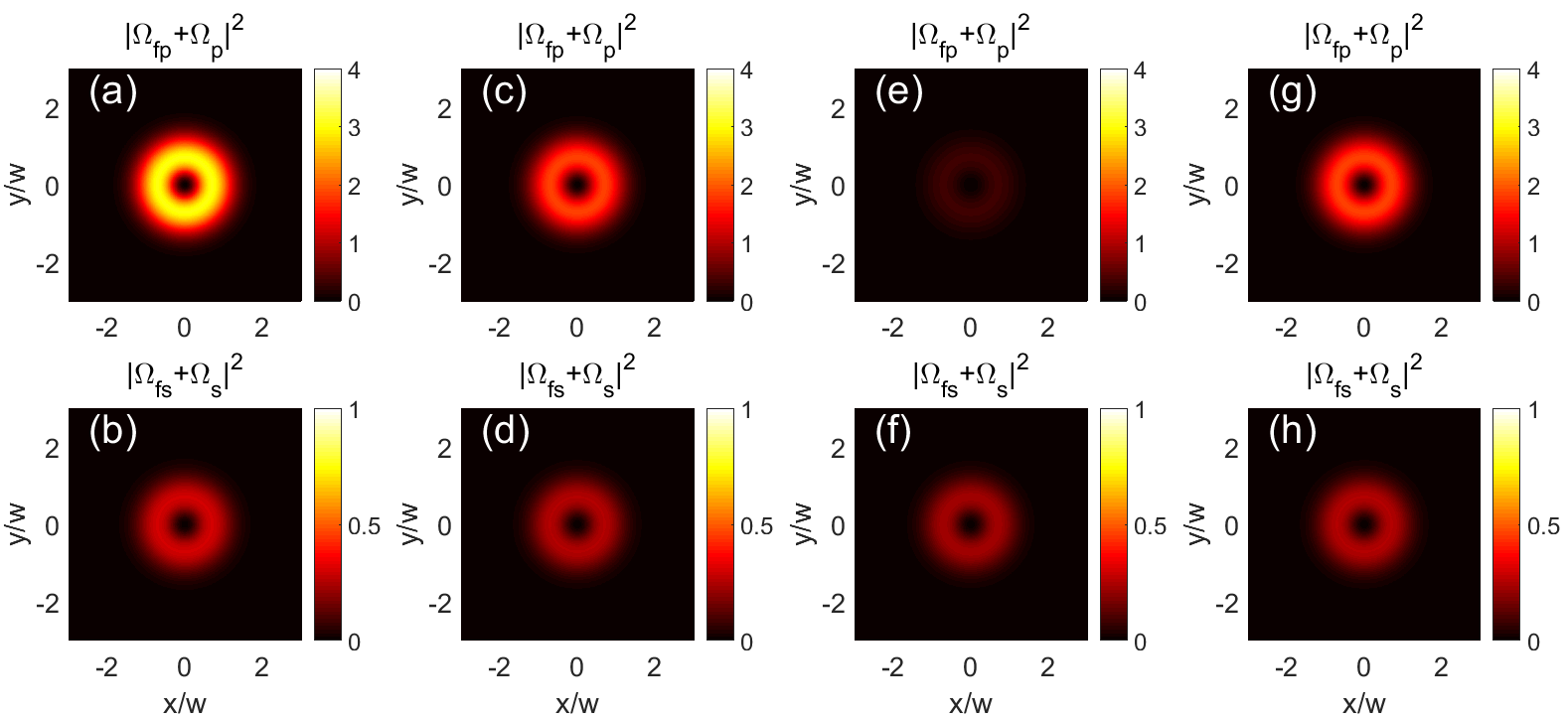}
\caption{The intensity distributions $\left |\Omega_{fp}+\Omega_p \right |^2/\gamma^2$ of the total output field with relative phase $\phi=\pi/2$ in (a), $\phi=\pi$ in (c), $\phi=3\pi/2$ in (e) and $\phi=2\pi$ in (g). 
The intensity distribution $\left |\Omega_{fs}+\Omega_s \right |^2/\gamma^2$ of the total output with relative phase $\phi=\pi/2$ in (b), $\phi=\pi$ in (d), $\phi=3\pi/2$ in (f) and $\phi=2\pi$ in (h). 
Here $l_p=l_s=1$, and the other parameters are the same as in Fig.~\ref{fig2}.}
\label{fig3}
\end{figure*}

\begin{figure*}[t]
\includegraphics[width=0.7\textwidth]{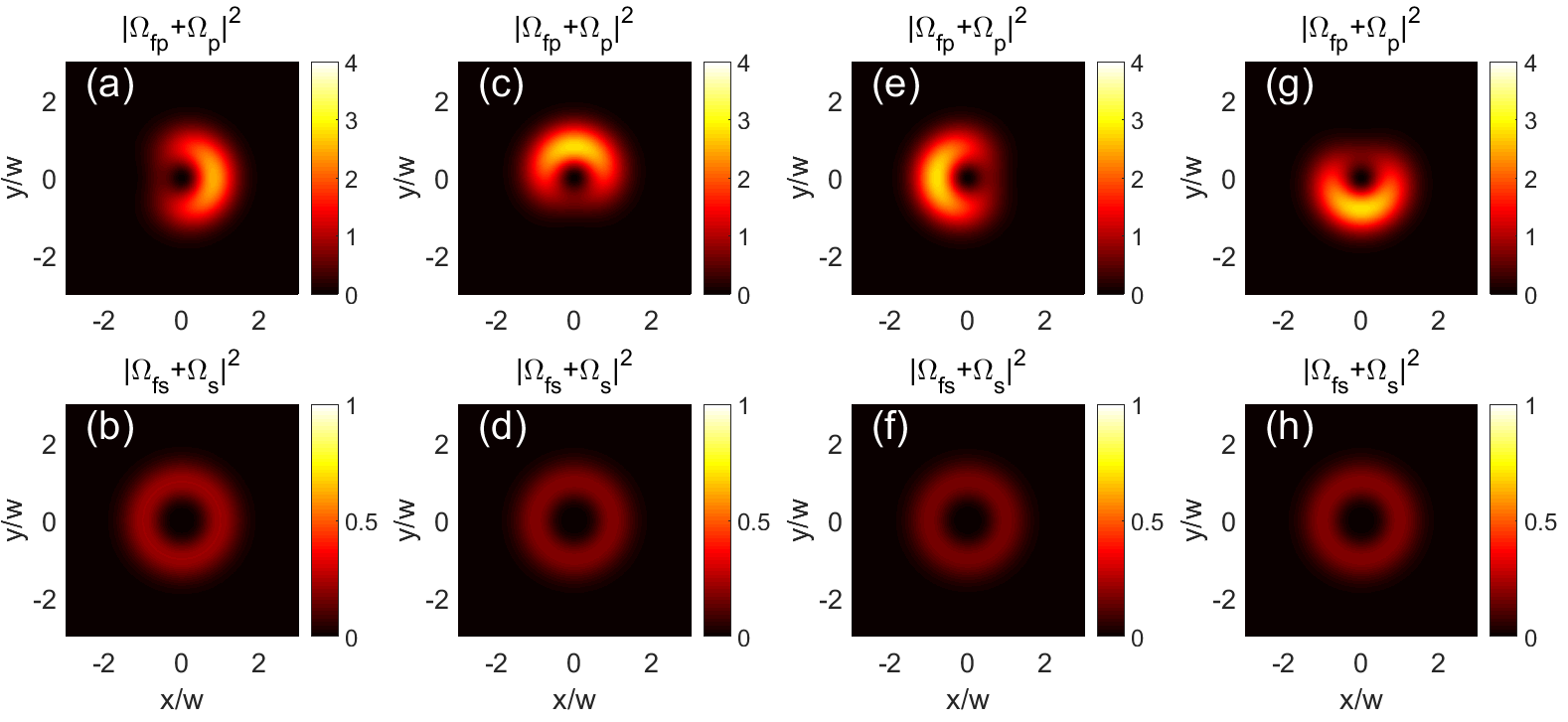}
\caption{The intensity distributions $\left |\Omega_{fp}+\Omega_p \right |^2/\gamma^2$ of the total output field with relative phase $\phi=\pi/2$ in (a), $\phi=\pi$ in (c), $\phi=3\pi/2$ in (e) and $\phi=2\pi$ in (g). 
The intensity distribution $\left |\Omega_{fs}+\Omega_s \right |^2/\gamma^2$ of the total output with relative phase $\phi=\pi/2$ in (b), $\phi=\pi$ in (d), $\phi=3\pi/2$ in (f) and $\phi=2\pi$ in (h).
Here $l_p=l_s=2$, and the other parameters are the same as in Fig.~\ref{fig2}.} 
\label{fig5}
\end{figure*}

Introducing a control field indeed enhances conversion efficiency; however, to explore the conversion mechanism, we simultaneously input two weak vortex optical fields to generate two FWM fields and examine the variations in the generated fields. 
When both the phase matching and resonance conditions are met, two pathways yielding the sum and difference frequency fields coexist, rendering the two light fields indistinguishable in terms of frequency, thereby allowing the total output fields $\Omega_{p,s}$ to be regarded as coherent superpositions of the two beams. 
Consequently, the interference between these two beams significantly affects the amplification and reduction of the output light. Equation (\ref{eq22}) reveals that the  OAM of the sum frequency field $\Omega_{fp}$ is controlled by the probe field $\Omega_{s}$, while the OAM of the difference frequency field originates from the field$ \Omega_{s}$.

From Fig.~\ref{fig3} (a,c,e,g), it is found that when the OAMs of the two probe beams are identical, the total intensity $|\Omega_p+\Omega_{fp}|^2$ of the combined output field $(\Omega_p+\Omega_{fp})$ exhibits sensitivity to the relative phase $\phi$ of the four fields, demonstrating periodic variations. 
At a relative phase of $\phi=\pi/2$, the total output field $(\Omega_p+\Omega_{fp})$ reaches its maximum intensity owing to constructive interference between the input field $\Omega_p$ and the generated field $\Omega_{fp}$. 
Conversely, when $\phi=\pi$, the total output field intensity $(\Omega_p+\Omega_{fp})$ reaches its minimum due to destructive interference between $\Omega_p$ and $\Omega_{fp}$. 
However, the total intensity $|\Omega_s+\Omega_{fs}|^2$ of the total output field $(\Omega_s+\Omega_{fs})$ remains robust against changes in the relative phase $\phi$. 
This stems from the fact that the intensity of the generated field $\Omega_{fp}$ is significantly smaller than that of $\Omega_{fs}$ due to the difference in the decoherence rates of the transitions $|4 \rangle \leftrightarrow |1 \rangle$ and $|2 \rangle \leftrightarrow |1 \rangle$. 
Consequently, no discernible optical interference occurs between $\Omega_s$ and $\Omega_{fs}$. 
Therefore, simultaneous asymmetric modulation of two total vortex beams can be achieved within such a system.



When two probe fields carrying different OAMs, as depicted in Fig. \ref{fig5} (a,c,e,g), the observation of the dimensionless intensity of the total output field $(\Omega_p+\Omega_{fp})$ illustrates that the interference of the two vortex beams results in a \textit{crescent-shaped} intensity distribution, with a distinct angle at which the amplitude is either suppressed or enhanced. 
Furthermore, it is intriguing that the intensity distribution of the total output vortex optical field undergoes rotation in the plane defined by axes $x$ and $y$ as the relative phase $\phi$ increases. 
This phenomenon signifies that phase modulation can alter the position of the intensity distribution due to changes in the interference pattern. 
Conversely, as shown in Fig. \ref{fig5} (b,d,f,h), the output field $(\Omega_s+\Omega_{fs})$ lacks these distinctive features since phase variation has minimal influence on its amplitude.



\section{conclusion}
We explore the propagation characteristics of OAM in vortex beams within a four-level cascade-type atomic system driven by two strong control fields and two weak probe fields. 
When either the probe or control fields carry OAM, the two FWM processes can generate sum-frequency and difference-frequency vortex light, inheriting the OAM of the input fields while inducing a change in optical frequency. 
Under resonance conditions, the generated field exhibits a transparent window around the vortex core due to the presence of a non-vortex control field. This ensures EIT during propagation and minimizes absorption losses \cite{qiangxishou,qiangxishou2}. 
Furthermore, we demonstrate the amplification and attenuation of the output vortex beam by phase adjustment using the dual-channel four-wave mixing process and beam interference. It is found that the phase sensitivities of the intensity distributions of the two output beams are very different.  
Adjusting the relative phase can allow for the spatial rotation of the intensity distribution of the output vortex beam. 
Such a modulation can be obviously observed when constructing the spatial structure, such as the crescent-shaped vortex beam.


\end{document}